\documentclass[modern, letterpaper]{aastex61}
\usepackage{hyperref}
\definecolor{linkcolor}{rgb}{0,0,1}
\hypersetup{colorlinks=true,linkcolor=linkcolor,citecolor=linkcolor,
            filecolor=linkcolor,urlcolor=linkcolor}
\hypersetup{pageanchor=true}
\usepackage{upgreek} 

\newcommand{\Disk}{{\rm disk}}
\newcommand{\Halo}{{\rm halo}}
\newcommand{\Barr}{{\rm bar}}

\usepackage{amsmath}
\usepackage{amsfonts}
\usepackage{amssymb}
\usepackage{booktabs}

\usepackage{graphicx}
\usepackage{color}

\newcommand{\documentname}{\textsl{Article}}
\newcommand{\sectionname}{Section}

\newcommand{\package}[1]{\texttt{#1}}

\newcommand{\python}{\package{Python}}

\newcommand{\project}[1]{\textsl{#1}}

\newcommand{\msun}{\ensuremath{\mathrm{M}_\odot}}
\newcommand{\kms}{\ensuremath{\mathrm{km}~\mathrm{s}^{-1}}}
\newcommand{\pc}{\ensuremath{\mathrm{pc}}}
\newcommand{\kpc}{\ensuremath{\mathrm{kpc}}}
\newcommand{\kmskpc}{\ensuremath{\mathrm{km}~\mathrm{s}^{-1}~\mathrm{kpc}^{-1}}}

\newcommand{\bs}[1]{\boldsymbol{#1}}

\shorttitle{Pal 5 and the Galactic Bar}
\shortauthors{PEARSON ET AL.}

\begin{document}
\title{Gaps and length asymmetry in the stellar stream Palomar 5 as effects of Galactic bar rotation}

\author{Sarah Pearson$^*$}
\affiliation{Department of Astronomy,
             Columbia University,
             550 W 120th St.,
             New York, NY 10027, USA}
\email{spearson@astro.columbia.edu}
\correspondingauthor{Sarah Pearson}

\author{Adrian~M.~Price-Whelan}
\affiliation{Department of Astrophysical Sciences,
             Princeton University, Princeton, NJ 08544, USA}

\author{Kathryn V. Johnston}
\affiliation{Department of Astronomy,
             Columbia University,
             550 W 120th St.,
             New York, NY 10027, USA}

\begin{abstract}
Recent Pan-STARRS data show that the leading arm from the globular cluster Palomar 5 (Pal 5) appears shorter than the trailing arm, while simulations of Pal 5 predict similar angular extents. We demonstrate that including the spinning Galactic bar with appropriate pattern speeds in the dynamical modeling of Pal 5 can reproduce the Pan-STARRS data. As the bar sweeps by, some stream stars experience a difference in net torques near pericenter. This leads to the formation of apparent gaps along Pal 5's tidal streams and these gaps grow due to an energy offset from the rest of the stream members. We conclude that only streams orbiting far from the Galactic center or streams on retrograde orbits (with respect to the bar) can be used to unambiguously constrain dark matter subhalo interactions. Additionally, we expect that the Pal 5 leading arm debris should re-appear south of the Pan-STARRS density truncation.
\end{abstract}

\keywords{dark matter --- Galaxy: structure --- Galaxy: kinematics and dynamics --- globular clusters: individual (Palomar 5) --- methods: numerical}

\section*{Introduction} \label{sec:intro}
The dense stellar stream emerging from the globular cluster Palomar 5 (Pal 5) is one of the few Galactic streams that is clearly associated with its progenitor system. The Pal 5 stream has therefore received much attention since its initial discovery (\citealt{oden01}) in the Sloan Digital Sky Survey (SDSS; \citealt{york00}). The SDSS photometric density maps of the stream, combined with subsequent kinematic measurements for both the cluster (\citealt{oden02,fritz15,dotter11}) and stream stars (\citealt{oden09,kuzma15}), have since enabled precise dynamical modeling of the density distribution along the Pal 5 stream to constrain the shape and radial profile of the mass distribution of the Milky Way (see, e.g., \citealt{kuepper15,bovy16}).

As seen by the SDSS and other northern follow-up (\citealt{ibata16}), the trailing arm extends $\approx$$15^\circ$ before fading into the background, whereas survey footprints have until recently limited exploration of the extent of the leading arm. The density structure of the trailing arm is not smooth: apparent density fluctuations along the trailing arm could indicate recent interactions with dark matter subhalos (e.g., \citealt{yoon11,carlberg12, erkal16, bovy17}), but it is still debated whether they arise from observational or dynamical effects (\citealt{thomas16}).

All previous simulations of the stream predict symmetric leading and trailing arms with similar angular extents at Pal 5's present day position (e.g., \citealt{dehnen04,pearson15,kuepper15}). It was therefore expected that the leading arm should extend several more degrees towards southern declinations. Recently-released photometric catalogs from the Pan-STARRS 1 survey (PS1; \citealt{chambers16}) enhance the southern coverage of the stream (\citealt{bernard16}), however these new data show that the leading arm only extends $\approx$$8^\circ$ from the cluster (progenitor) center before abruptly ending. It is unlikely that this apparent truncation is caused by
observational selection effects from matched filtering because the stream extension is not seen in nearer or farther distance bins (\citealt{bernard16}), so what has cut off the leading arm of Pal 5?

Most previous studies have modeled the evolution of Pal 5 in an analytic, static Milky Way potential consisting of a bulge, disk, and dark matter halo. However, recent work has demonstrated that including a time-dependent, triaxial bar can greatly affect the morphologies of streams (\citealt{hattori16,price16b}). It has also been shown that the Galactic bar could create density differences between leading and trailing arm of Pal 5 (\citealt{erkal16}), even though Pal 5's perigalacticon ($\approx$7--8 kpc) is far from the supposed extent of the bar ($\approx$4 kpc; e.g., \citealt{wegg13}). Motivated by these theoretical and observational findings, in this \documentname\ we further explore the affect of a rotating Galactic bar on the morphology of the Pal 5 stream.

\section{Methods}\label{sec:methods}

\subsection{Potential}\label{sec:pot}
\begin{figure*}
\centerline{\includegraphics[width=\textwidth]{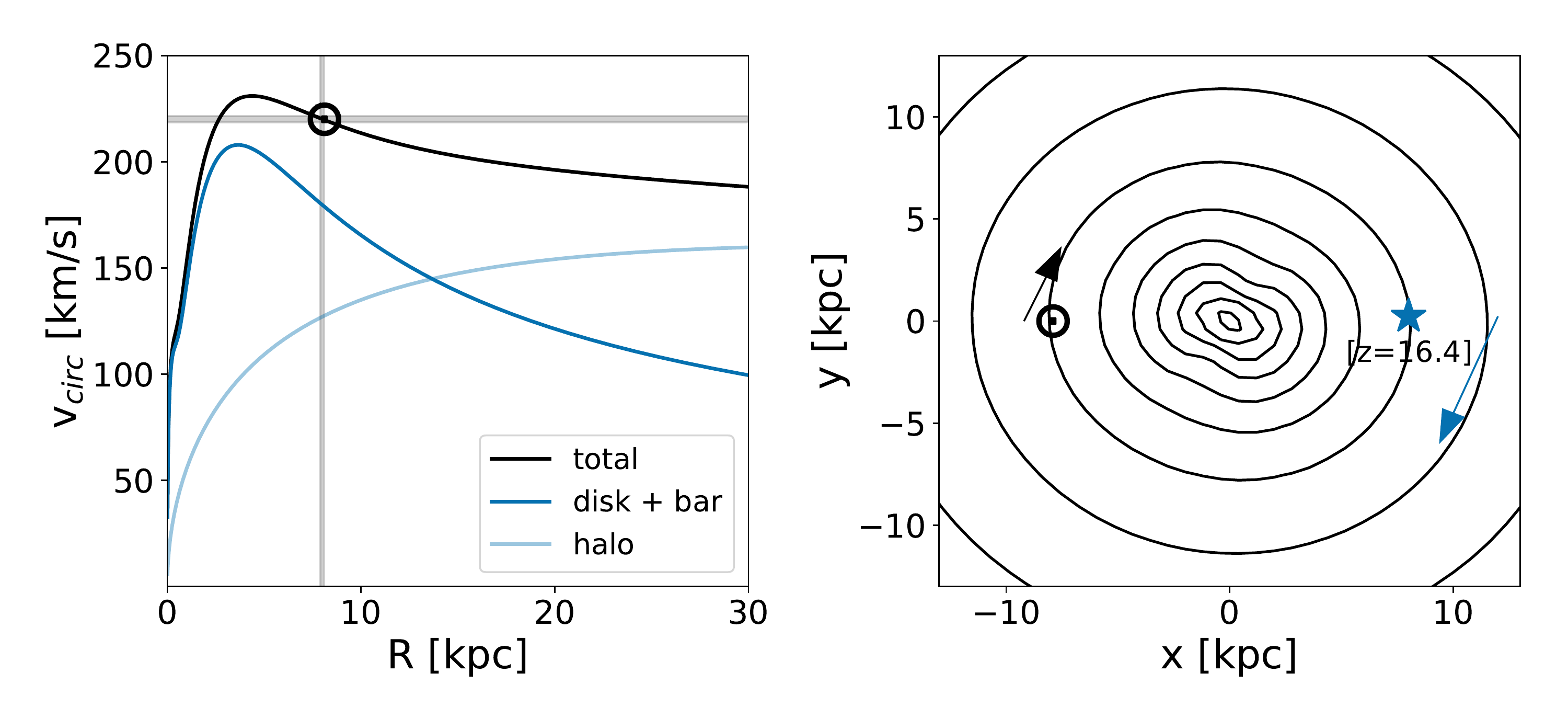}}
\caption{Left: Circular  velocity  curves for the barred Milky Way potential model introduced in Section \ref{sec:pot}. The black  line  shows  the sum of all  components,  the lines  below  show a decomposition by potential components.  Vertical and horizontal gray lines shows the approximate position of the Sun ($\odot$) and its circular velocity. Right: Contours of constant surface density in the plane for the barred Milky Way potential at present day (i.e. $\alpha = 27^\circ$). $\odot$ indicates the position of the Sun, and the blue star indicates Pal 5's projected position in the plane of the Galaxy (it is located 16.4 kpc above the plane). In this projection, the direction of motion is clockwise for the Sun and Pal 5 (see arrows).}
\label{fig:pot}
\end{figure*}

We use a three-component mass model to represent the gravitational field of the
Milky Way with disk, halo, and bar components as shown in the left panel of Figure \ref{fig:pot}. In detail, we use a
Miyamoto-Nagai disk for the disk (\citealt{miya75}) and a flattened
Navarro-Frenk-White profile for the halo (NFW; \citealt{navarro96}).
Expressed as gravitational potentials, the disk and halo models are parametrized
as:
\begin{eqnarray}
\Phi_\Disk(R,z) &=& -\frac{G \, M_\Disk}
    {\sqrt{R^2 + \left(a_\Disk + \sqrt{z^2 + b_\Disk^2}\right)^2}} \\
\Phi_\Halo(R,z) &=& -\frac{G \, M_\Halo}{r_\Halo} \, \frac{\ln(u + 1)}{u}
\end{eqnarray}
where $R$ is the cylindrical radius, $z$ is the standard Cartesian coordinate,
and
\begin{eqnarray}
u &=& \frac{\sqrt{R^2+z^2/q_\Halo^2}}{r_\Halo} \quad .
\end{eqnarray}
All other variables are parameters set to values described in \tablename~\ref{tbl:potential-params}.
The bar potential is a low-order basis-function expansion (BFE) representation of a triaxial, exponential
density profile (\citealt{wang12,price16b}) given by
\begin{eqnarray}\label{eq:4}
\rho_\Barr &=& \rho_0 \left[ \exp\left(-r_1^2/2\right) + r_2^{-1.85}\,\exp(-r_2) \right]\\
r_1 &=& \left[ \left((x/x_0)^2 + (y/y_0)^2 \right)^2 + (z/z_0)^4 \right]^{1/4}\label{eq:5}\\
r_2 &=& \left[ \frac{q^2\,(x^2+y^2) + z^2}{z_0^2} \right]^{1/2}\label{eq:6}
\end{eqnarray}
with scale-lengths fixed to $x_0=1.49~\kpc$, $y_0=0.58~\kpc$,
$z_0=0.4~\kpc$, and $q=0.6$ (\citealt{dwek95,wang12}). Following previous work, we use the ``self-consistent field''  BFE formalism (SCF; \citealt{hernquist92}). The physical scales of the bar are reflected in the  scale-mass and scale-radius assumed in the BFE (see Table 1), with the remaining radial and angular behavior of the bar in Equation \ref{eq:4}, \ref{eq:5} and \ref{eq:6} captured by including expansion terms up to n=2 and l=6. We do not include a spherically symmetric Galactic bulge because the mass in this component is likely small compared to the bar ($M_\Barr = 10^{10}~\msun$, \citealt{portail15}).

The bar component rotates with frequency vector anti-aligned with the $z$
coordinate axis $\boldsymbol{\Omega} = -\Omega_b \, \hat{z}$ (i.e. in the
direction of Galactic rotation) and has a present day (t=0) angular offset from the Galactic $x$-axis in the direction of rotation, $\alpha =  27^\circ$, which is consistent with
observations.
We set the disk and halo potential parameters following previous studies
(\citealt{bovy15,bovy16}), but remove mass from the disk to include the bar
component.
We have checked that the cylindrically-averaged rotation curve and surface
density profiles are consistent with recent measurements of these quantities
($v_{\rm circ,\odot} \approx 220~\kms$; \citealt{bovy12}, see Table \ref{tbl:potential-params} and Figure \ref{fig:pot}) and that the
enclosed mass profile at larger radii is consistent with constraints from halo
tracers (e.g., \citealt{xue08,deason12,kuepper15}).

\begin{floattable}
\begin{deluxetable}{r l}
\tabletypesize{\footnotesize}
\caption{Parameters for the experiment
\label{tbl:potential-params}}

\tablehead{%
    \colhead{name} & \colhead{value}
}
\startdata
$M_\Halo$ & $5 \times 10^{11}~\msun$ \\
$r_\Halo$ & $18~\kpc$ \\
$q_\Halo$ & $0.94$ \\
\hline
$M_\Disk$ & $6 \times 10^{10}~\msun$ \\
$R_\Disk$ & $3~\kpc$ \\
$z_\Disk$ & $280~\pc$ \\
\hline
$M_{{\rm bar},s}$& $10^{10}~\msun$$^{(a)}$ \\
$r_{{\rm bar},s}$ & $1.1~\kpc$ \\
$\alpha$ & $27~{\rm deg}$ \\
$\Omega_b$ & \textit{varied}\\
\hline
Pal 5 (RA, Dec) & ($229,-0.124)^\circ$$^{(b)}$ \\
$v_r$ & $-58.7 ~\kms$$^{(c)}$ \\
$d$ & 22.9 kpc \\
($\mu_{\alpha}cos\delta,\mu_{\delta}$) & (-2.296,-2.257) mas yr$^{-1}$$^{(d)}$\\
\enddata
\footnote{\citealt{portail15}}
\footnote{\citealt{oden02}} 
\footnote{\citealt{bovy16}}
\footnote{\citealt{fritz15}}
\end{deluxetable}

\end{floattable}

\subsection{Orbit integration and mock stream generation}\label{sec:stream-method}

We use a C-implementation of the Dormand-Prince 8th-order Runge-Kutta scheme \citep{prince81,hairer93} to integrate orbits in the above potential (wrapped in \python\ and released with \package{gala}; \citealt{gala-v0.1.3}). We use a time-step of $\Delta t = 0.5~{\rm Myr}$, which conserves (Jacobi) energy with $|\Delta E_J/E_{J,0}| \leq 10^{-11}$ over our longest integration periods.

We use the ``particle spray'' method (see \citealt{fardal15}) implemented in \package{gala} (\citealt{gala-v0.1.3}) to generate model stellar streams: star particles are created near the Lagrange points of a progenitor system on a given orbit with dispersions in position and velocity set by the mass and orbit of the progenitor (see  Section 2.4 in \citealt{fardal15}). 
In this work, we do not include the gravitational influence of the progenitor system on particles after released; this will affect the detailed density distribution of the simulated stream (\citealt{gibbons14}) but we do not expect it to affect qualitative comparisons of the stream morphology. We set the initial cluster mass to $m = 50000~\msun$ and release two particles (one at each Lagrange point) every time step. 

\subsection{Experiments}\label{sec:exp}

To investigate the influence of the Galactic bar on the Pal 5 stream, we fix the Galactic potential parameters (see Section \ref{sec:pot}) and the 6D phase space coordinates of the Pal 5 cluster (see Table \ref{tbl:potential-params}). 
We use \package{astropy} (\citealt{astropy13}) and \package{gala} (\citealt{gala-v0.1.3}) to transform these coordinates to a Galactocentric frame assuming the Sun is at Galactocentric position $(x,y,z) = (-8.0,0,0)~{\rm kpc}$ \citep[e.g.,][]{schoenrich12} with velocity $(v_x,v_y,v_z) = (-11.1, 244, 7.25)~\kms$ \citep[e.g.,][]{schoenrich10, schoenrich12}.

The bar spin in the Milky Way is prograde with respect to the disk and Pal 5's orbit around the Galaxy, and measurements of the bar pattern speed  span from $\approx$25--$70~\kmskpc$ (\citealt{gerhard11}). However, to explore the full effect of adding a bar to the Galactic potential, we investigate three different scenarios:
\begin{itemize}
\item[1.] Non-rotating, static bar, $\Omega_{b}=0$ km s$^{-1}$ kpc$^{-1}$.
\item[2.] Retrograde bar, $\Omega_{b}=  -60$ km s$^{-1}$ kpc$^{-1}$.
\item[3.] Prograde bar, $\Omega_{b}= 20$ to $80$ km s$^{-1}$ kpc$^{-1}$, with 1 km s$^{-1}$ kpc$^{-1}$ increments.
\end{itemize}
For each choice of pattern speed, we generate mock stellar streams following the prescription described above (\sectionname~\ref{sec:stream-method}): we first integrate the progenitor orbit backwards in the time-dependent Milky Way model from Pal 5's present day position for 8000 timesteps ($4~{\rm Gyr}$), then begin the stream-generating procedure, integrating the orbit and all stream particles and the bar forward to present-day.


\section{Results}\label{sec:results}
\subsection{Streams generated with a prograde bar can reproduce the PS1 truncation}\label{sec:hole}
\begin{figure*}
\centerline{\includegraphics[width=\textwidth]{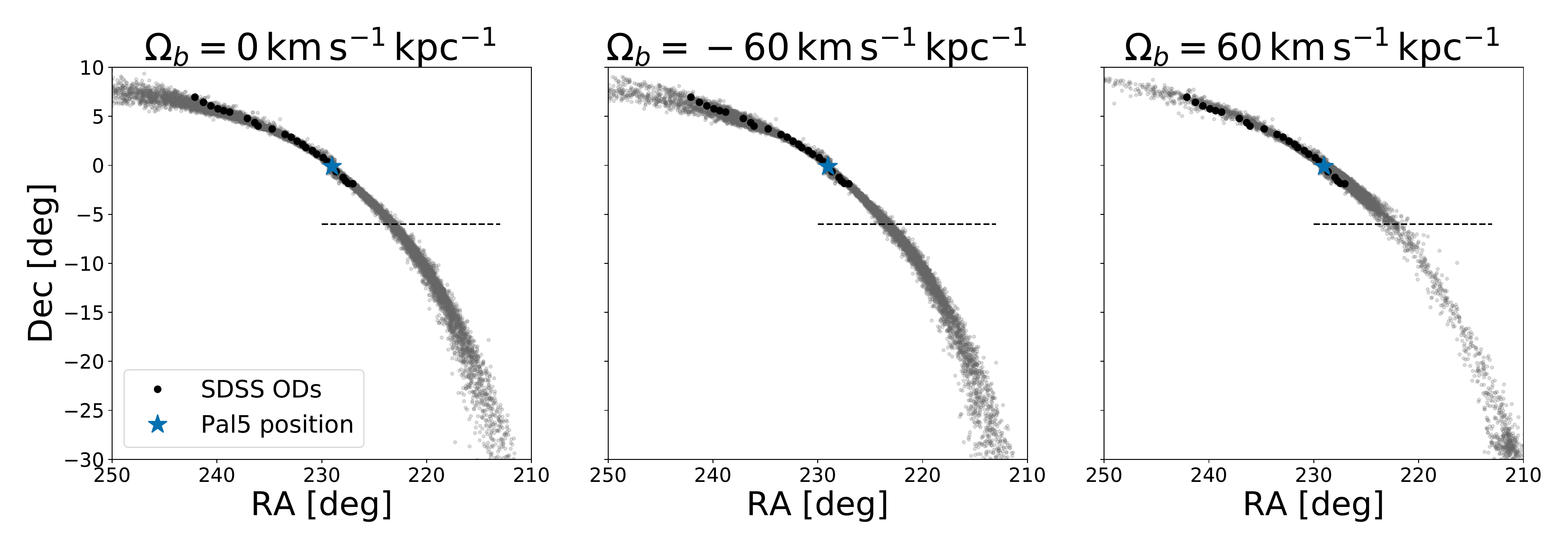}}
\caption{Sky projection of mock Pal 5 stream particles (grey) evolved in a potential with a static bar (left), retrograde bar with $\Omega_b = -60~\kmskpc$ (middle), and a prograde bar with $\Omega_b = 60~\kmskpc$ (right). Over-plotted are the SDSS photometric over-density locations (black). The dashed line demonstrates where the Pal 5 stream appears to end in the PS1 data (\citealt{bernard16}); this observed density decline of the leading arm is qualitatively reproduced with the prograde bar (right panel). The PS1 footprint extends to $\delta = -30^\circ$, the lower limit of these panels.}
\label{fig:gc_hel}
\end{figure*}

As expected, including a static bar (\figurename~\ref{fig:gc_hel}, left panel) or a retrograde bar (\figurename~\ref{fig:gc_hel}, middle panel) does not much change the properties of the model stream (see also \citealt{erkal16}). For both of these cases, the stream curvature is qualitatively reproduced---compare model stars (grey points) to SDSS over-density positions (black points)---and the leading and trailing arm both extend symmetrically $\approx$$15^\circ$ in sky projection.

In contrast, the mock streams generated in a Galactic potential with a prograde bar generically, lead to apparent gaps (caused by an under density) and length differences along both the leading and trailing arm. From the grid of pattern speeds (\sectionname~\ref{sec:exp}), we select streams that qualitatively match the curvature of the trailing arm in sky position and the radial velocity gradient along the stream (e.g., \citealt{oden09, kuzma15}). From these, we then select model streams that display abrupt density drops in the leading arm. In particular, for the mock stream generated with a bar pattern speed $\Omega_{b} = 60~\kmskpc$ (\figurename~\ref{fig:gc_hel}, right panel), we see that the dense portion of the trailing arm extends $\approx$$15^\circ$, while the leading arm has a drastic density drop at the location of the Pan-STARRS leading arm truncation, $(\alpha,\delta)\approx(222, -6)^\circ$ (dashed line), and extends only $\approx$$8^\circ$ from the cluster center.

\subsection{Observational signatures}

Our models suggest that the truncation of the leading arm in the PS1 map of Pal 5 should be an apparent gap and not a complete cutoff of the stream: more stream stars could be located farther south along the stream trajectory, as suggested by the over-density of stars near $(\alpha,\delta)\approx (212,-29)^\circ$ seen in \figurename~\ref{fig:gc_hel}. Here we explore the observable phase-space morphology of these stars.

\begin{figure*}
\centerline{\includegraphics[width=\textwidth]{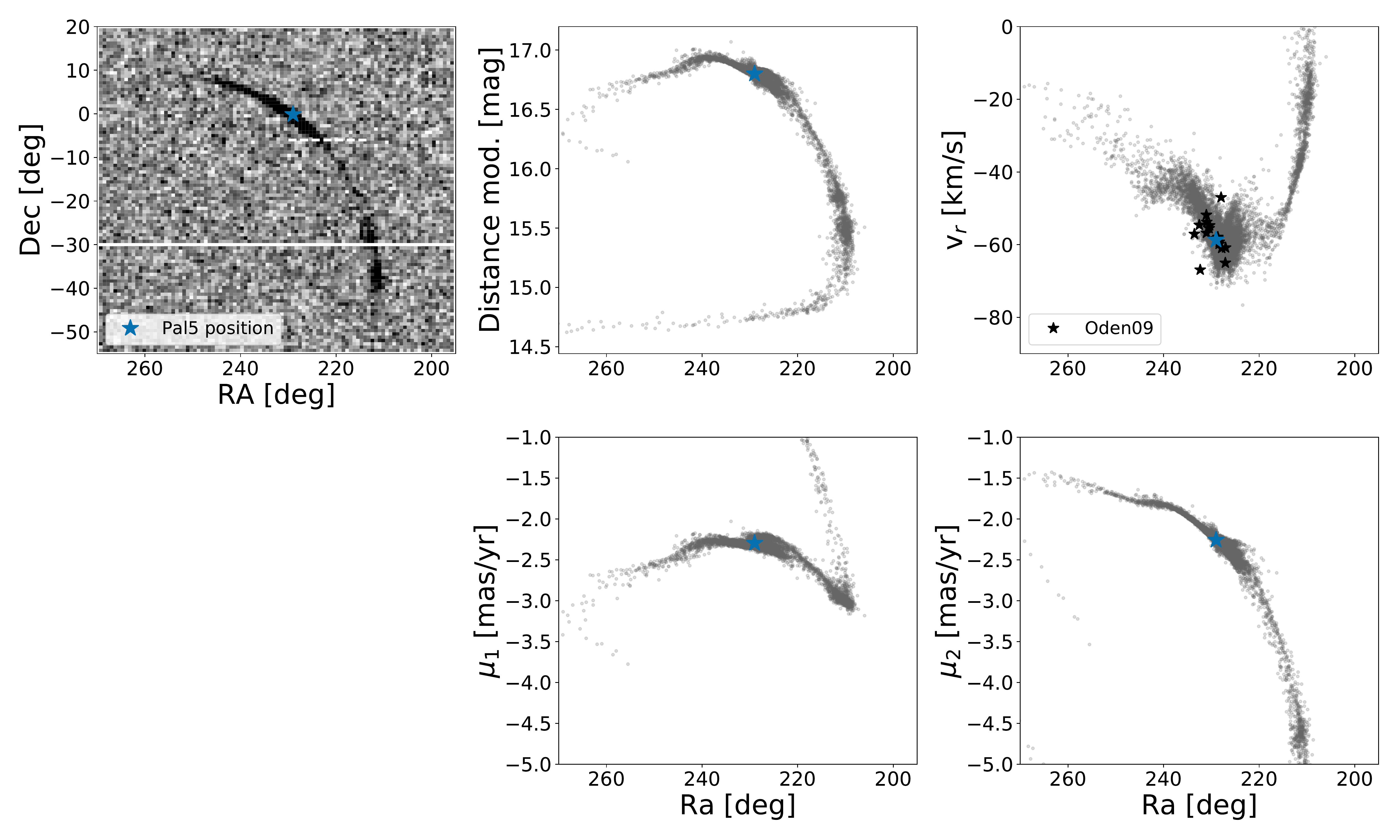}}
\caption{Left: simulated stellar number-count map of the model stream (\citealt{balbinot17}). Right: observables of our Pal 5 mock stream evolved in a potential with $\Omega_b = 60 ~\kmskpc$. We expect that Pal 5's leading arm should reappear south of the reported Pan-STARRS truncation (dashed line, upper left panel). The solid line shows the lower limit of the PS1 footprint. Based on the predicted distance modulus, radial velocity and proper motions, we encourage observers to look for the rest of Pal 5's leading arm stars.}
\label{fig:debris}
\end{figure*}

\figurename~\ref{fig:debris} (top left panel) shows a simulated stellar number-count map of the model stream with a uniform background of stars with mean density $0.058~{\rm stars}~{\rm arcmin}^{-2}$ (\citealt{balbinot17}); this panel shows more clearly that a dense portion of the leading arm of the mock stream generated with a prograde bar reappears near the edge of the PS1 footprint (at $\delta = -30^\circ$). It is important to note that the particle spray method used (with uniform mass-loss) does not accurately reproduce the density along the stream. However, relative to the other dense portions of the stream, this stream model predicts that the dense clumps of particles around $(\alpha,\delta)\approx (212,-30)^\circ$ to $(212,-40)^\circ$ should have similar surface densities and therefore could be detected with extended imaging of this region. Other pattern speeds ($\Omega_b = 30, 41, 52, 62, 72, 79 ~\kmskpc$) that display density drops at similar locations also show similar morphologies. A common prediction from these model streams is that, if the stream has been disrupting for several Gigayears (as expected; \citealt{kuepper15}), dense stream debris should exist along the extended stream track in southern declinations. The exact location of gaps and further debris depends on the motion of Pal 5, the pattern speed of the bar, and the given potential (see Section \ref{sec:growth}).

These extensions of the stream would have unique kinematic properties relative to the unperturbed stream stars, as is shown in the other panels in \figurename~\ref{fig:debris} that plot the mock stream particles in other observable dimensions. First note that the distance modulus of these clumps is over 1 magnitude brighter, so a matched-filtering procedure using an isochrone at the cluster's distance would not find these stars. Also note the discrepant velocity structure between the near-cluster stream stars and these clumps. With data from the second data release of the \project{Gaia} mission (\citealt{gaia16}), proper motions for red-giant stars (known to exist in the stream; \citealt{kuzma15}) at these distance moduli will have uncertainties around $200~\mu{\rm as}~{\rm yr}^{-1}$ (estimated using \package{PyGaia}); using proper motions in combination with photometric matched-filtering should greatly enhance the contrast of the stream in this region and help test for the existence of this other associated debris.

Interestingly, our mock stream generated in the potential with $\Omega_b = 60~\kmskpc$, shows an apparent truncation of the trailing arm as well $(\alpha,\delta)\approx (242,8)^\circ$, which is also seen in the PS1 data.

\subsection{``Gap'' forming mechanism}\label{sec:mech}

The presence of a rotating, triaxial Galactic bar in the Milky Way potential model can create gaps and under-densities in simulated mock streams that are not observed when the bar orientation is fixed. The bar must therefore be asymmetrically perturbing or torquing particle orbits. We expect the most important perturbations to occur at orbital phases where a pericentric passage coincides with a crossing of the Galactic plane when the stream particles are typically closest to the bar. We also expect the magnitude of the perturbation to depend on the phase of the bar relative to a given particle's orbit. Those with pericenters ahead of the bar along their orbits will be pulled back by the bar, while those behind will be pulled ahead. Note that, in our case the angular momenta of both the bar and the orbit are negative, so pulling back or forward actually corresponds to a positive and negative torque respectively. The result of an encounter manifests as differences in net torques at adjacent points in the stream, which induce net differences in energy and therefore over time can evolve to form a gap. Hence, the gap will grow due to the energy offset from the rest of the stream members, simply as a tidal stream grows due to an energy offset from the released stars and their progenitor (\citealt{johnston98, johnston01, helmi99}). 

To isolate and illustrate this gap formation mechanism, we generate isotropic Gaussian-distributed balls of test particles around the Pal 5 progenitor orbit and follow the orbits of these particles. The advantage of using an isotropic ball of particles instead of the ``particle spray'' technique is that all particles begin to phase mix at a single time instead of being released at uniform time steps and this allows clear examination of the interaction sequence. We integrate the 6D position of the Pal 5 cluster backwards in time from its present-day position for 4000 Myr (from $t=0$ to $t=-4000~{\rm Myr}$). From the endpoint of the backwards integration, we generate initial conditions sampled from an isotropic Gaussian in position, with dispersion $\sigma_x = \left(\frac{m_{\rm Pal 5}}{M(<r)}\right)^{1/3} \, r $ where $M(<r)$ is the enclosed Milky Way mass and $r$ the instantaneous orbital radius. We assume ${m_{\rm Pal 5}} = 50000 ~\msun$, $r \approx 7-8$ kpc, which yields $M(<r) \approx 1.5 \times 10^{11}  ~\msun$ and $\sigma_x \approx 50 ~{\rm pc}$. Additionally we sample from an isotropic Gaussian in velocity, with dispersion $\sigma_v = 1 ~\kms$, comparable to the velocity  dispersion  of  the  cluster.
These dispersions will cause the ball particles to shear and will therefore approach the bar with slightly different phase angles each time they individually reach pericenter.

\begin{figure*}
\centerline{\includegraphics[width=\textwidth]{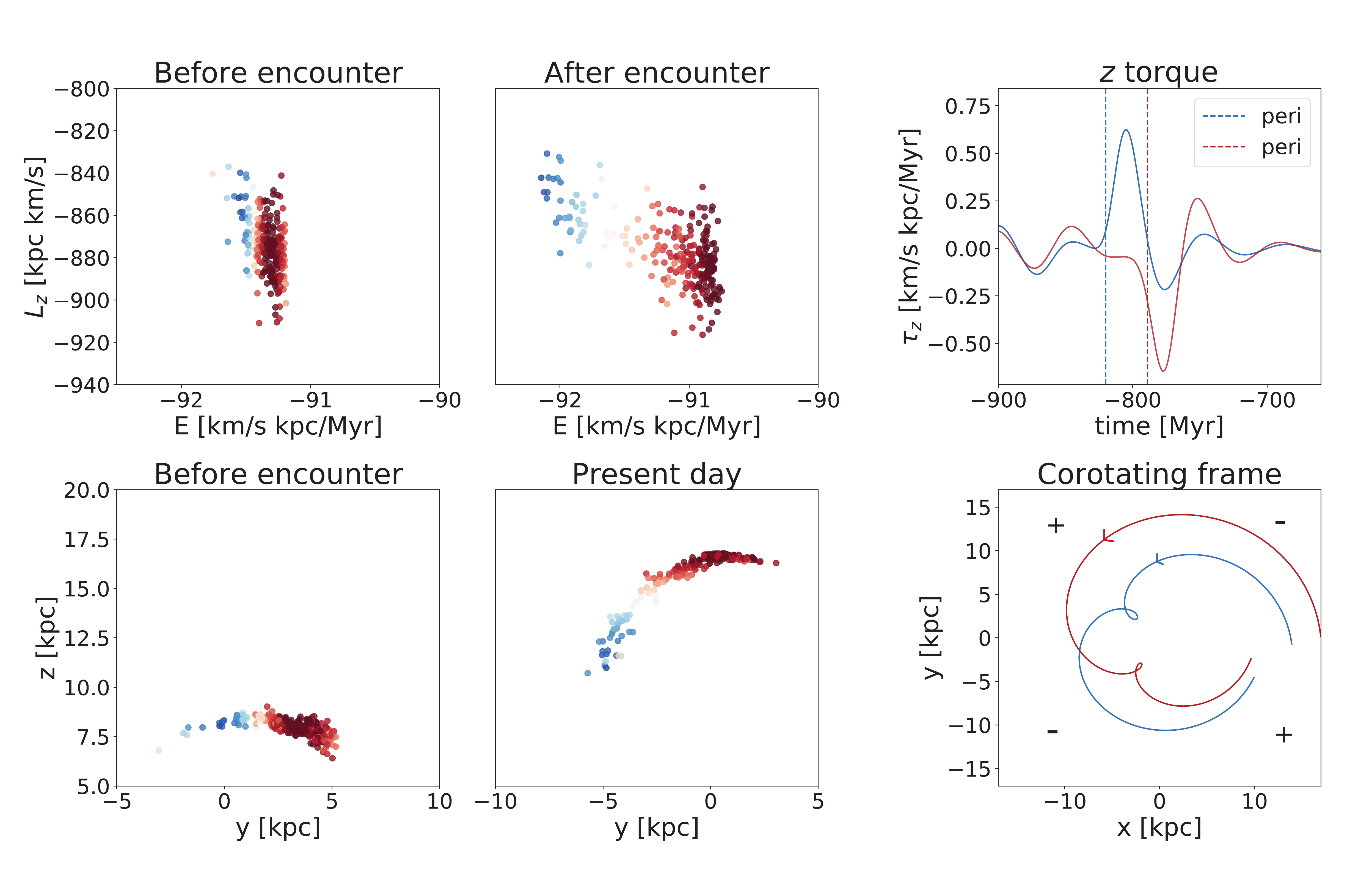}}
\caption{{\sl Left two columns}: time evolution of energy $E$ vs. $z$-component of angular momentum $L_z$ (top) and physical positions (bottom) of a ball of particles generated to roughly represent the leading arm of Pal 5 in a barred potential with $\Omega_b = 60~\kmskpc$. {\sl Right column, top}: particles that receive the largest net positive torque (blue) and largest net negative torque (red) near their orbital pericenter (dashed vertical lines, top). {\sl Right column, bottom}: Orbits of the same two particles near pericenter (${\rm t} = -900 {\rm ~Myr}$ to ${\rm t} = -700 {\rm ~Myr}$) in the corotating frame with the bar aligned with the horizontal axis. The arrows show the direction of the orbits and  ``+''/``-''  indicate the sign of the bar's torque in the corotating frame (see also \citealt{hattori16}, Figure 1). The blue particle experiences a positive torque as it reaches pericenter (see ``loop'') while the red particle, which reaches pericenter slightly later, receives a negative torque from the bar. This leads to a difference in the particles' net torque (i.e. $z$ angular momentum) and therefore an offset of the blue and red points in energy and physical space (middle column).}

\label{fig:holeform60}
\end{figure*}

We focus on a particular orbital pericenter of the mean orbit (at $t \approx -790~{\rm Myr}$). From the endpoint of the backwards integration, we first evolve the particles forward in a static bar potential from $t = -4000~{\rm Myr}$ to $t = -1000~{\rm Myr}$. We then turn on the bar with a pattern speed $\Omega_b = 60~\kmskpc$, evolve the particles and bar to $t_{\rm stop} = -700~{\rm Myr}$ where we again freeze the bar, and then continue integrating the orbits of the particles until present-day. \figurename~\ref{fig:holeform60} summarizes the results of this procedure. The leftmost column shows the star particles initially identified as ``leading'' particles (lower energy relative to the progenitor at $t=-4000~{\rm Myr}$) in the $z$ component of the angular momentum ($L_z$) vs energy ($E$) space (top panel) and in projected cartesian, Galactocentric coordinates (bottom panel), at ${\rm t} = -850 {\rm ~Myr}$ prior to a pericenter encounter with the bar. The middle column shows the same particles after the encounter with the bar at ${\rm t} = -700 {\rm ~Myr}$ (top). The particles are color coded based on their total shift in angular momentum (bluer particles experience larger positive shifts and redder particles experience larger negative shifts). It is clear that there have been changes in angular momenta and therefore the spread in energy increases dramatically after the bar encounter. 
Comparing the left column panels to the middle column panels, it is evident that the blue and red particles move away from each other (during the encounter) in energy space (top) and over time also in physical space (see t = 0, bottom) leading to the formation of an under-density in real-space as the particles phase-mix.

Pal 5-like orbits have a large inclination relative to the Galactic midplane ($z=0$), therefore the ball particles that are spread in phase approximately around some mean orbit will reach pericenter and cross the midplane at different times, when the bar is at different angles within the plane.
\figurename~\ref{fig:holeform60} (right column) shows the two particles that experience the largest net positive (blue) and negative (red) torque in this pericenter encounter (top), and their position with respect to the bar in the corotating bar frame where the orbit directions are counter clockwise (see arrows bottom, right panel). As expected, the two particles see the bar in a different orientation as they reach pericenter (see ``loops'', bottom panel). The overall effect is to decrease/increase the magnitude of their angular momenta (as seen in the evolution of $L_z$ from left to middle panel) which corresponds to a positive and negative torque respectively (see +/- in \figurename~\ref{fig:holeform60}, bottom right).

A prograde bar can lead to differences in net torques because of the difference in position of the stars with respect to the bar at their orbital pericenter. Due to the high inclination of Pal 5's orbit, the Pal 5 gaps form and grow because of individual encounters with the bar which differs from the resonant affect of the bar on streams orbiting in the Galactic plane (\citealt{hattori16}).\footnote{See \citet{hattori16} for example of how streams orbiting in the Galactic plane can be shortened/lengthened if the leading vs trailing stream periodically approach pericenter at different phases with respect to the bar.} Stars will not experience this interaction with the bar in a potential with a static bar. With a retrograde bar, the effect is not as important because the interaction time is greatly reduced.

\subsection{``Gap'' growth}\label{sec:growth}

While the energy and angular momentum of each particle is not conserved in the time dependent barred potential, the Jacobi Energy ($E_J = E - \bs{\Omega} \cdot \bs{L}$) is conserved for each particle (\citealt{binney08}). We can therefore calculate the offset in energy for each particle experiencing a bar encounter from the change in their $z$ angular momentum:
\begin{align}
\Delta E_J &= \Delta E - \bs{\Omega} \cdot \Delta \bs{L} = 0
\end{align}
\begin{align}
\Delta E = -\Omega_b \, \Delta L_z \quad .
\end{align}
The change in the particle's $z$ angular momentum can be expressed as the integral of the torque over a given interaction with the bar that begins and ends at $(t_-, t_+)$ or a sum over $k$ torques computed over the above interval at $K$ times with fixed timestep $\Delta t$:
\begin{equation}
\Delta L_z = \int_{t_-}^{t_+} \tau_z \, dt \approx \Delta t \, \sum_k^K \tau_{z,k}\quad ,
\end{equation}
The particles that experience a net positive or negative torque will be offset from the rest of the particles in energy and will therefore have different orbital times. This small difference in azimuthal orbital periods, $\Delta T_\Psi$, will then cause an angular separation for each particle, $w$, that grows by an amount per orbital period:
\begin{equation}\label{eq:w}
\Delta w \approx 2 \pi \, \frac{\Delta T_\Psi}{T_\Psi} \quad .
\end{equation}
In general, the orbital period depends strongly on $E$ and only weakly on $L_z$ (see \citet{hendel15}, \figurename~2), therefore we assume that the azimuthal orbital period is similar for a particle on a circular orbit with the same energy as a particle on an eccentric orbit. We can represent the flat part of the rotation curve (see Figure \ref{fig:pot}) as a logarithmic potential with circular velocity, $v_c \approx 200~\kms$,  and can therefore express $\frac{\Delta T_\Psi}{T_\Psi} = \frac{\Delta E}{v_c^2}$. We can then express the angular offset for one particle experiencing a net torque as:
\begin{equation}\label{eq:w}
\Delta w = 2 \pi \, \frac{\Delta E}{v_c^2} = 2 \pi \, \frac{-\Omega_b \, \Delta L_{z}}{v_c^2} \quad .
\end{equation}
As some particles are positively torqued and some are negatively torqued in a given bar encounter, we can express the apparent gap growth, $\Lambda$, per orbital time based on the this torque difference, $\xi$: 
\begin{equation}\label{eq:w}
\Lambda \approx 2 \pi\, \frac{\Omega_b}{v_c^2} \, \left(\Delta L_{z, {\rm pos}} - \Delta L_{z, {\rm neg}}\right) = 2 \pi\, \frac{\Omega_b}{v_c^2} ~\xi \quad .
\end{equation}
From the example shown in the upper right panel of \figurename~\ref{fig:holeform60}, we find that $\Delta L_{z,pos} = 7.7~\kms ~{\rm kpc}$ and $\Delta L_{z,neg} = -10.7 ~\kms ~{\rm kpc}$. Hence $\xi = 18.4 ~\kms ~{\rm kpc}$, so given the assumptions stated above we can express the gap growth per orbital time as:

\begin{equation}\label{eq:l}
\Lambda = 10^{\circ} \times \frac{\Omega_b}{60 ~\kmskpc}  \times \frac{(200 ~\kms)^2}{v_c^2}  \times \frac{\xi}{18.4 ~\kms ~{\rm kpc}}  \quad .
\end{equation}
Recall that we here used the most net negatively and net positively torqued particles, and that the actual size of the apparent gap will grow to roughly half of this ($\approx$ $5 ^{\circ}$, see \ref{fig:holeform60} top, middle panel) within an orbital time in this specific example.

\section{Discussion}\label{sec:disc}

\subsection{Repeated encounters}\label{sec:rep}

\citet{hattori16} showed that the long-term influence of a time-dependent bar can affect the length of streams orbiting in the plane of the galaxy through repeated, resonant torques from the bar. This phenomenon occurs as stream stars periodically
encounter the bar near pericenter; if the stars repeatedly approach the bar with the same or similar phase with respect to the bar, they experience the so-called ``shepherding'' effect from resonant bar encounters (\citealt{hattori16}).

Our analysis differs from the resonant picture described in \citet{hattori16} as Pal 5 is not orbiting in the plane of the Galaxy. The Pal 5 gaps form and grow because of individual encounters with stars close to their orbital pericenters. Because of the high inclination of the Pal 5 orbit, different particles can be affected at each approach towards pericenter so that repeated, resonant encounters do not occur.

\begin{figure*}
\centerline{\includegraphics[width=\textwidth]{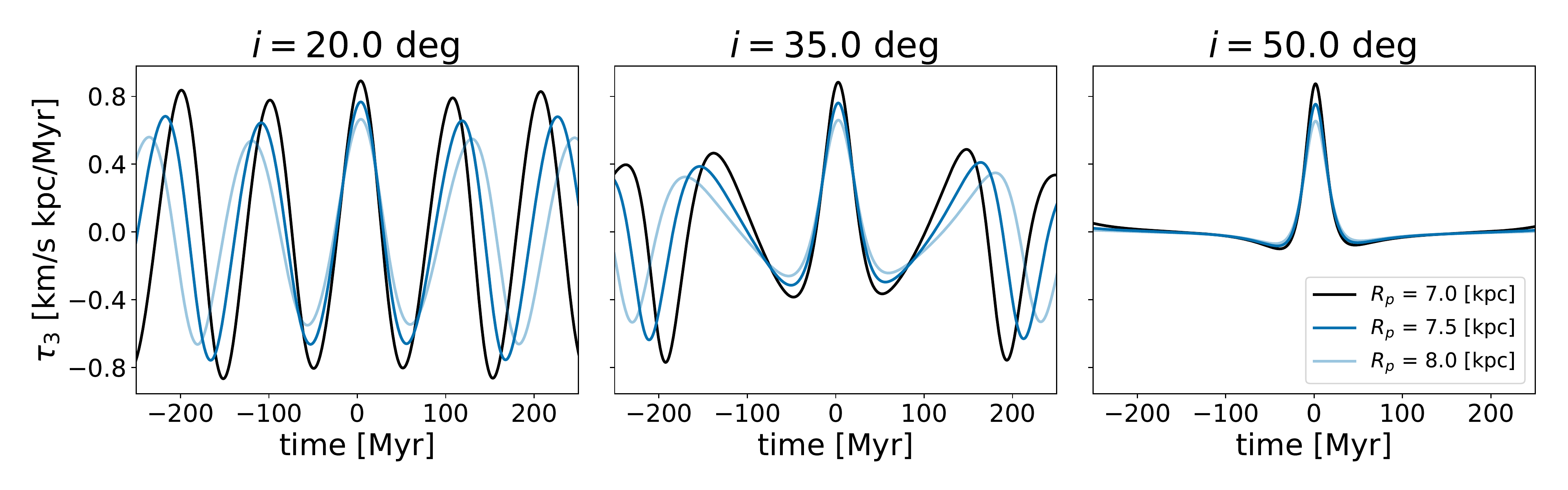}}
\caption{Particles with three different pericentric distances (lines), experience a pericenter and bar passage at the midplane at $t=0$. Small $R_p$ yields a larger magnitude in the z-direction of the torque. Small orbital inclinations (left), induce periodic bar encounters, while large inclinations (right) are dominated by one bar encounter event, when the particle passes the midplane.}
\label{fig:rperi}
\end{figure*}

In \figurename~\ref{fig:rperi}, we explore the effect of varying the inclination with respect to the bar plane, $i = (20, 35, 50)^\circ$, for a single particle orbiting the Galaxy with three different pericentric distances, $R_p = (7, 7.5, 8)~\kpc$. In each case, a pericenter and bar passage occurs at $t=0$ at the midplane (where the torque from the bar is maximum). When the inclination is small (left panel), the particle repeatedly experiences the bar sweeping past it (i.e. a torque in the z-direction). At larger inclinations (middle and right panels), the bar is most important as a particle passes the midplane and can lead to net torques even from single encounters. As expected, a smaller pericentric distance will yield a larger magnitude of the torque (lines ordered by brightness). Additionally, the net torque (and the size of the apparent gap) will depend on the interaction time which is set by the details of the specific orbit.

At high inclinations, the location along the stream where the net torque perturbs particles will change over each orbit. This will lead to more stochastic perturbations and gap formation that can drastically alter or ``wash out'' the appearance of under-densities from previous encounters. The net affect can therefore, in principle, lead to complicated stream density structures with apparent gaps of any size. Additionally, since the gap growth depends on the difference in net torque that particles receive after a bar encounter, arbitrarily small gaps can form with varying inclination and encounter timescale (in contrast with the idea that the bar can only create large-scale stream asymmetries; \citealt{erkal16}).

\subsection{Consequence for subhalo search}

According to the $\Lambda$-Cold Dark Matter ($\Lambda$CDM) cosmological model, our galaxy is expected to be filled with thousands of dark matter subhalos, in contrast with the number of galactic satellites ($\approx$100s) observed around the Milky Way. Recently, alternative dark matter theories have been proposed that suppress small-scale clustering to make the predicted number of subhalos smaller (e.g., ``fuzzy'' dark matter; \citealt{hui16}). Including baryons in cosmological CDM simulations also seems to suppress small-scale structure; the number of subhalos within $<25$ kpc of the Galactic center for a simulated, Milky Way-like galaxy is reduced by a factor of $\approx$5 because of interactions with the stellar disk (e.g., \citealt{kimmel17}).

One way to test these theories is to find evidence for ``dark'' subhalos that may have thus far remained undetected because they are only indirectly observable (\citealt{klypin99}). One such way to infer the presence of dark subhalos is to search for recent interactions between subhalos and cold stellar streams that leave density perturbations in the form of apparent gaps. Several groups have shown that interactions with subhalos, in abundances similar to those expected from CDM simulations, can induce observable density variations along the Pal 5 stream (e.g., \citealt{yoon11,carlberg12,erkal16,bovy17}).

Our work demonstrates an alternative explanation for density variations in stellar streams: interactions with the Galactic bar. From the experiments above, this seems to happen for streams that orbit prograde with respect to the Galactic bar and may only mimic subhalo interactions when the orbital inclinations are large. We therefore conclude that distant cold streams and streams on retrograde orbits with respect to the disk (such as GD1) will be least affected by bar encounters and therefore most suitable for potential indirect dark matter subhalo detections. 

\section{Conclusion}\label{sec:conc}
We have demonstrated that the Galactic bar can create apparent gaps (under-densities) in the Pal 5 stream as the bar sweeps past stream stars as they reach their orbital pericenter.
In particular we found that:
\begin{enumerate}
\item The length asymmetry between the leading and trailing arms of Pal 5 seen in PS1 can be reproduced by including a (prograde) bar in a Galactic potential model consistent with current observational constraints for the Milky Way and Pal 5's orbit.
\item A generic expectation of our models is the reappearance of debris south of the PS1 truncation.  While the exact location depends on Pal 5's orbit, the bar parameters and the Milky Way potential, observers should search for the debris to test this hypothesis.
\item Under-densities form when particles experience different net torques from bar encounters that depend on their phase with the bar at pericenter. The apparent gaps grow in time because of induced energy offsets from the rest of the stream members, similar to
how gaps grow after encounters with subhalos or how streams grow due to an energy offset from the released stars and their progenitor.
\item Dark matter subhalos which encounter streams can also lead to the formation of gaps. However, these signatures will be ambiguous in stellar streams because of similar signatures induced by the bar. We should therefore search for subhalo encounters in streams with large peri-centers or streams on retrograde orbits.
\end{enumerate}
We confirm that including the bar when modeling streams in the inner Milky Way is critical (e.g., \citealt{price16b}, \citealt{hattori16}). In future work, we investigate whether all detected under densities in the leading and trailing arms of Pal 5 (e.g. \citealt{thomas16}, \citealt{erkal16}) can be solely attributed to the bar, by doing a rigorous comparison of bar gap signatures to other gap signatures (see e.g. \citealt{amorisco16, sanders16, erkal16a, bovy17}, Sandford et al., {\it submitted}) in an Nbody simulation of the Pal 5 stream.


\acknowledgements
We are pleased to acknowledge Anthony Brown and the Gaia Project Scientist
Support Team and the Gaia Data Processing and Analysis Consortium (DPAC) for
making the \package{PyGaia} package open-source. We thank the Flatiron Institute Center for Computational Astrophysics for providing the space to carry out this project. SP thanks Julio Chanam\'e for insightful discussions. KVJ and  SP acknowledge support from NSF grant AST-1614743.

\end{document}